# Investigation of artificial domains realised by local gallium focused ion beam (FIB) modification of Pt/Co/Pt trilayer structures

A. Aziz, S. J. Bending, H. Roberts, and S. Crampin

Department of Physics, University of Bath, Claverton Down, Bath BA2 7AY, UK

P. J. Heard

Interface Analysis Centre, University of Bristol, Bristol BS2 8BS, UK

C. H. Marrows

School of Physics and Astronomy, University of Leeds, Leeds LS2 9JT, UK

We present the results of experimental investigations of magnetic switching and magnetotransport in a new generation of magnetic devices containing artificially patterned domains. Our devices are realised by locally reducing the coercive field of a perpendicularly magnetised Pt (3.5 nm)/Co (0.5 nm)/Pt (1.6 nm) trilayer structure using a gallium focused ion beam (FIB). Artificial domain walls are created at the interfaces between dosed and undosed regions when an external magnetic field switches the former but not the latter. We have exploited this property to create stripe-like domains with widths down to sub-micron lengthscales, separated by undosed regions. Using the extraordinary Hall effect to monitor the local magnetisation we have investigated the reversal dynamics of these artificial domains by measuring major and minor hysteresis loops. The coercive field of regions irradiated with identical doses systematically increases as their size decreases. In the lower branch of minor loops, reversal is seen to occur via a few large Barkhausen events. Preliminary measurements of transport across domain walls reveal a positive domain wall resistance, that does not change sign from 4.2 K to 300 K.

7215Gd, 7425Ha, 7570-i, 6116Ch

**Section I: Introduction**

It is well established that the magnetisation of a very thin Co film sandwiched between Pt layers, prefers to align perpendicular to the plane due to surface and interface anisotropies[1]. The shape and position of the magnetic domains, which form to reduce the magnetic dipole energy of the film, can be difficult to predict and tend to change with each magnetic cycle, and such materials are poorly suited for devices which exploit domain wall (DW) resistance effects. DW locations can be controlled, to a limited extent, by fabricating notches, zigzag structures and constrictions in the magnetic films[2,3,4], yet the actual shapes and sizes of domains remain determined by intrinsic film parameters like surface, shape, interface and crystalline anisotropies. Full control of the properties of magnetic domains can, however, be achieved provided local anisotropies are changed in a systematic fashion. This can be realised in perpendicularly magnetised Pt/Co/Pt systems by local irradiation with a gallium focused ion beam (FIB), which reduces the film coercivity due to relaxation and mixing of the interfaces[5]. Using Ga FIB irradiation it is now possible to *define* artificial static magnetic domains of arbitrary shapes and sizes in Pt/Co/Pt sandwich or multilayer structures with deep sub-micron precision. Unlike natural domains, artificial domains are fabricated by engineering the intrinsic spin properties of the multilayers and promise reproducible magnetic (and magnetotransport) properties. In this manuscript we attempt to demonstrate the strong potential of this approach. In section III, the magnetic switching and area scaling of the dosed regions are studied using the extraordinary Hall effect (EHE). The coercive field of regions irradiated with identical doses systematically increases as their size decreases. In the lower branch of minor loops, reversal is seen to occur via a few large Barkhausen events.

Artificial domains may also prove to be an important tool for studying spin transport properties *across* domain walls and we describe a preliminary investigation of DW resistance in a lateral magnetic 'superlattice' in section IV.

**Section II: Device fabrication procedure**

A Pt(3.5 nm)/Co(0.5 nm)/Pt(1.6 nm) sandwich structure was deposited at 300 K using dc magnetron sputtering with an Ar pressure of 2.7 mTorr. The very thin (0.5 nm) Co layer used in our experiment is probably not *chemically* continuous, and will contain some pin holes. However, since the Pt develops an induced magnetic moment in the proximity of Co[6], it will be *magnetically* continuous. The RMS roughness of the deposited film was 0.4 nm, as measured by AFM. The magnetic films were then patterned into Hall bar structures, based on the intersection of 2 μm wide wires, using optical lithography and reactive ion etching with a 1:1 mixture of $SF_6$ and Ar. Unusually thin Pt capping layers have been used in these experiments to minimise the effect of current shunting out of the Co film. This renders the structures extremely sensitive to the FIB, and an additional 8 nm $SiO_2$ 'attenuation' layer has been deposited on the top surface prior to irradiation. This distributes the recoil energy more uniformly to the Pt/Co interface through collision cascades. Focused ion beam irradiation is performed on the completed structure with a commercial FIB (FEI Strata 201) at an incident Ga ion energy of 30 keV and a 1 pA current. The beam diameter was about 10 nm, the distance between neighboring pixels 7.4 nm, and irradiation was performed at a magnification of 10,000×, yielding a write field of $30.4 \times 28.5$ μm$^2$. An ion dose of 0.007 pC/μm$^2$ is achieved, using pixel dwell times of 0.4 μs. The properties of the gallium ions after transmission through the $SiO_2$ layer have been estimated using SRIM (The Stopping and Range of Ions in Matter) software[7] and the average energy (% of Ga ions transmitted) of the gallium ions after passing through 8 nm $SiO_2$ are estimated to be 19.6 keV (99.6 %).

**Section III: Properties of artificial domains**

The switching properties of perpendicularly magnetised dosed regions were investigated using the extraordinary Hall effect (EHE)[8]. Rectangular regions of x × 3 $\mu m^2$ (where x=250 nm, 500 nm, 1μm, 2 μm and 4 μm) were lightly dosed (0.007 pC/$\mu m^2$) using FIB in the middle of a Hall cross. A sketch of a typical device is shown in the inset of Fig. 1, where the AC current flows (I=10 μA, 29 Hz) between contacts on the left and right, and L, M and R are labels for three adjacent Hall crosses. Figure 1 shows the EHE voltage measured at the middle Hall cross (M) (containing a 1 μm wide dosed stripe) which exhibits three steps. Changes in EHE voltage are directly related to the out-of-plane component of the magnetisation of the magnetic film in the Hall cross region. Step A corresponds to the switching of the dosed region whereas steps B and C are due to the undosed regions on the left and right sides. The minor loops 1 and 2 (gray lines) are obtained after saturating the sample in positive and negative magnetic fields respectively. These are shown in more detail in Figure 2 which illustrates how the form of minor loops of identically dosed regions scales with their area after saturation at positive fields (the sweep rate for all minor loops was 2.6 Oe/s). The width of the dosed region was varied from 500 nm to 4 μm while the length was kept constant at 3 μm. (A 250 nm wide sample was also measured but its switching field lay very close to that of the undosed region making it difficult to measure a minor loop.) The following four features are clearly identifiable in Figure 2. (a) Minor loops of dosed regions of width smaller than 2 μm are asymmetric and (b) the asymmetry increases with the decreasing width of the dosed regions. (c) For data in the lower half of the figure, when H is swept in a positive sense, switching occurs at around 20 Oe, i.e. it is nearly independent of the width of the dosed region. (d) Discrete Barkhausen steps are also observable in the lower half of the minor loops.

The asymmetry of the minor loops shown in Figure 2a indicates that the switching mechanism is different when the magnetic field is varied in a positive or negative sense. When all regions (dosed and undosed) are magnetised in the same direction (upper half), dosed regions switch sharply and the switching field increases strongly as the irradiated area decreases, as shown graphically in fig 2b. It shows the logical trend that when the area approaches zero, $H_c^d / H_c^{ud} \rightarrow 1$. Since this switching field is greater than the switching fields observed for the lower halves of the minor loops, it seems probable that reversal is governed by domain nucleation followed by domain wall motion. The increased switching field for small domains (<2 μm) may be due to the decrease in number of available nucleation sites. When an external magnetic field is applied in this way, such that it switches the dosed but not the undosed region, a magnetic domain will form whose size is determined by the irradiated region, with a domain wall at its border. We observe that switching back to a uniform magnetisation from this state (lower half) does not depend strongly on the width of the dosed region, suggesting that reversal is now controlled by DW motion alone.

In addition to the switching mechanism, minor loops also shed light on the local pinning potentials in the dosed region. Switching in the lower half of the minor loops is not as sharp as in the upper half and has many plateaux-like features, which can most clearly be seen in the 1 μm dosed sample. When the external field is increased from negative values up to zero, a gradual change in EHE voltage is observed in the narrower structures indicating that magnetisation of the irradiated region is already beginning to relax. Using the Hall effect to quantify this we estimate that areas of ~0.1 μm², ~0.5 μm², ~0.7 μm² and ~0.7 μm² are relaxed for 4μm, 2μm, 1μm and 0.5μm wide regions respectively at H=0 (i.e. domains with width <4 μm have significantly less than 100% remanent in these minor loops).

For positively increasing fields the reversal of the 1μm wide irradiated stripe (area 3 μm²) occurs predominantly in six approximately equal sized steps, which can be associated with local pinning sites present in the Ga ion dosed region. We estimate that an area A≈0.4 μm² reverses in each step, which must be correlated to large scale interface irregularities. For a Co film of thickness 0.5 nm, the magnetisation volume reversed during each step is $2 \times 10^{-16}$ cm³ and, assuming that interface irregularities are distributed uniformly, the characteristic separation of strong pinning sites is estimated as $\sqrt{A} \approx 600 nm$ for the Ga ion irradiated region. This is fairly consistent with the observation that artificial domains of width ≤ 500 nm generally reverse in a single step after the relaxation phase (H<0).

**Section IV: Preliminary domain wall transport measurements**

Many experimental studies of domain wall resistance have been described in the literature, but the area remains somewhat controversial; even the issue of the actual sign of DW resistance (positive or negative) is not yet fully resolved [9]. Apparently DW resistance is usually very small and can be masked by spurious effects like anisotropy magnetoresistance (AMR), Lorentz magnetoresistance (LMR) and EHE. Magnetoresistance is also present in perpendicularly magnetised systems when closure domains exist near the surface[10]. LMR is usually small, but for the systems where $2\pi f_c \tau \approx 1$, where $f_c$ is the cyclotron frequency and $\tau$ is the average relaxation time, it can be significant at low temperatures[11]. Measurements on artificial domains, of the type discussed here, promise to overcome all the above mentioned problems. Closure domains are not energetically favored in ultrathin (0.5 nm) Co films with strong perpendicular anisotropy ($Q=K/2\pi M_s \gg 1$), and the associated AMR effect is absent. Furthermore, because of the high film resistivity (31 μΩ cm and 26 μΩ cm at 300 K and 4.2 K respectively) $\omega_c\tau \sim 10^{-4} \ll 1$ and LMR and any MR effect associated with the wiggling of current lines at domain walls can be neglected at all temperatures.

We have made preliminary measurements of domain wall resistance in Wheatstone bridge structures like the one illustrated in Fig. 3. A 1μm period lateral 'superlattice' of six artificial stripe domains (DW perpendicular to direction of current flow) was patterned in two asymmetric arms of the bridge in order to maximise the magnitude of the measured signal. Magnetoresistance measurements in the range 4.2 K-300 K show a positive change which can be associated with the formation of domain walls at the interfaces between dosed and undosed regions, consistent with the theoretical spin mixing model[12]. One advantage of the use of artificial domains is that their direction with respect to the that of the current can be varied at will, making them an ideal test bed for theoretical models of domain wall scattering. A detailed investigation of DW scattering in such structures is in progress and will be presented in a future publication.

**Conclusion:**

We have described the magnetic properties of a new generation of magnetic devices in which artificial magnetic domains are introduced into perpendicularly magnetised Pt (3.5 nm)/Co (0.5 nm)/Pt (1.6 nm) trilayer structures by local Ga FIB irradiation. The switching properties and size-scaling of the dosed regions has been studied using EHE, allowing probable reversal mechanisms to be established. Discrete steps are observed in the minor loops of dosed regions which allow a direct estimate of the Barkhausen volume. Artificial DWs are created at the border of dosed and undosed regions when an external magnetic field switches the former but not the latter, and this property has been used to create stripe-like domains. Preliminary measurements of DW resistance using this approach indicate a positive value over the entire temperature range 4.2 K-300 K which is in fair agreement with theoretical predictions[12].

We acknowledge financial support for this work from the Leverhulme Trust through research project grant *F/00 351/F*

[1] W. B. Zeper, J. A. M. Greidanus, P. F. Carcia, and C. R. Fincher, J. Appl. Phys. **65**, 4971 (1989).

[2] M. Kläui, C. A. F. Vaz, A. Lapicki, T. Suzuki, Z. Cui, and J. A. C. Bland, Microelectron. Eng. **73-74**, 785 (2004).

[3] T. Taniyama, I. Nakatani, T. Namikawa, and Y. Yamazaki, Phys. Rev. Lett. **82**, 2780 (1999).

[4] S. Lepadatu and Y. B. Xu, Phys. Rev. Lett. **92**, 127201 (2004).

[5] C. Vieu, J. Gierak, H. Launois, T. Aign, P. Meyer, J. P. Jamet, J. Ferré, C. Chappert, T. Devolder, V. Mathet, and H. Bernas, J. Appl. Phys. **91**, 3103 (2002).

[6] S. Stähler, G. Schütz, P. Fischer, M. Knülle, S. Rüegg, S. Parkin, H. Ebert, and W. B. Zeper, J. Magn. Magn. Mater. **121**, 234 (1993).

[7] J. F. Zielgler, J. Biersack, and U. Littmark, computer software SRIM (2003). Available from www.srim.org.

[8] C. L. Canedy, X. W. Li, and G. Xiao, J. Appl. Phys. **81**, 5367 (1997).

[9] A. D. Kent, J. Yu, U. Rüdiger, and S. S. P. Parkin, J. Phys.: Condens. Matter **13**, R461 (2001).

[10] I. Knittel, S. Faas, M. A. Gothe, M. R. Koblischka, and U. Hartmann, J. Magn. Magn. Mater. **272-276**, e1431 (2004).

[11] U. Rüdiger, J. Yu, S. S. P. Parkin, and A. D. Kent, J. Magn. Magn. Mater. **198-199**, 261 (1999).

[12] P. M. Levy and S. Zhang, Phys. Rev. Lett. **25**, 5110 (1997).

**Figure Captions**

FIG. 1: EHE voltage measurements as a function of perpendicular magnetic field at 300 K at a Hall cross (M) which has a 1 μm x 3 μm dosed region at its centre. The inset shows a sketch of the device. L, M and R are three adjacent Hall crosses. The central parts of the left (L) and right (R) Hall crosses are undosed. The light rectangular region at the centre of the middle cross (M) is the irradiated region.

FIG. 2: (a) EHE voltage measurements at 300 K of minor loops of irradiated regions with widths varying from 500 nm up to 4 μm. Samples were first saturated in positive fields, and the external field then swept down until the dosed region reversed, at which point it was swept back up again. (b) The fractional change in the coercive field of the dosed region ($H_c^d$) with respect to the average coercive field of the undosed regions either side ($H_c^{ud}$) as a function of the area of the dosed region.

FIG. 3: (a) Optical micrograph of the Wheatstone bridge structure used to investigate domain wall resistance. A 10 μA AC current flows between A and B, and the voltage is measured between the leads $V_1$ and $V_2$. The expanded region shows the top two Pt(1.6 nm)/Co(0.5 nm)/Pt(3.5 nm) leads of dimension 1 μm × 16 μm. Thick 150 nm Ti/Au fingers are deposited to create low resistance voltage contacts to the Pt/Co/Pt wires. Six dosed regions of width 1μm separated

**by 1µm undosed regions are patterned in the top left and the bottom right Pt/Co/Pt leads. (b) Schematic of the top leads of the bridge structure.**

FIG. 1

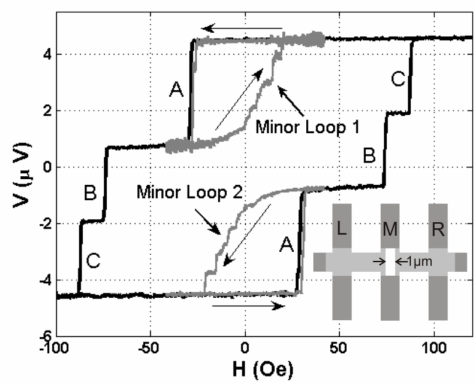
12

FIG. 2:

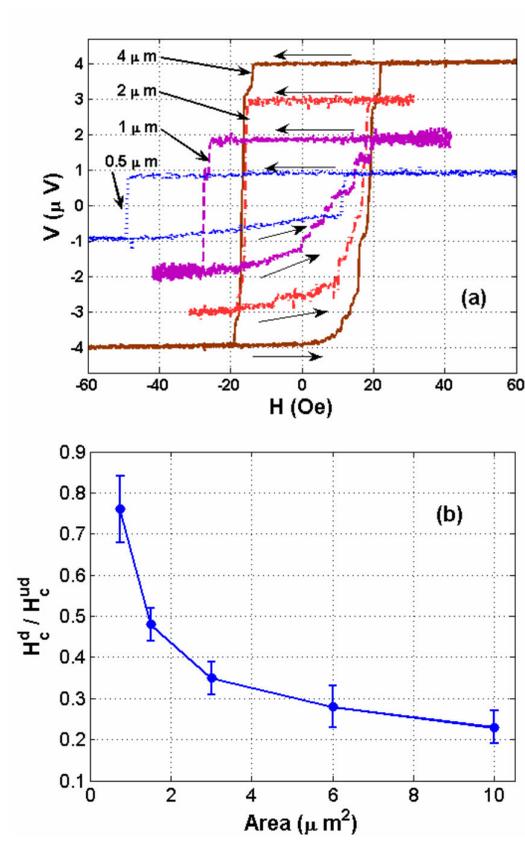

FIG. 3:

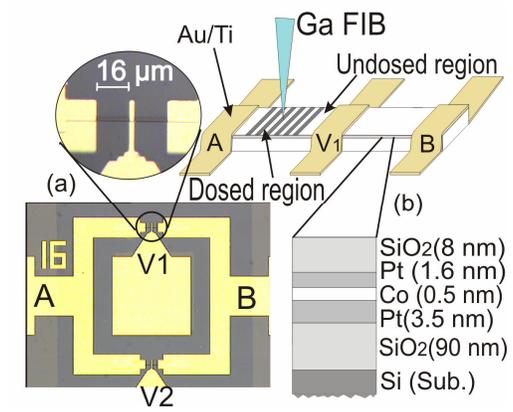